\documentclass[prb,twocolumn]{revtex4}

\usepackage{amsmath,amsfonts,amssymb,bm}
\usepackage{dcolumn}
\usepackage[final]{graphicx}
\usepackage{bm}
\usepackage{times}

\newcommand*{\E}[0]{\mathcal{E}}
\renewcommand*{\P}[0]{\mathcal{P}}

\begin{document}
\bibliographystyle{apsrev}

\title{Four-wave mixing in coupled semiconductor quantum dots}

\author{Pekka Koskinen}\thanks{Permanent address: Department of Physics, University of Jyv\=askyl\=a, 40014 Jyv\=askyl\=a, Finland}
\author{Ulrich Hohenester}\email{ulrich.hohenester@uni-graz.at}
\affiliation{Institut f\"ur Theoretische Physik,
  Karl--Franzens--Universit\"at Graz, Universit\"atsplatz 5,
  8010 Graz, Austria}

\date{June 7, 2002}

\begin{abstract}

We present a theoretical analysis of four-wave mixing in coupled quantum dots subject to inhomogeneous broadening. For the biexciton transitions a clear signature of interdot-coupling appears in the spectra. The possibility of experimental observation is discussed.

\end{abstract}


\maketitle

Few-particle states in optically excited semiconductor quantum dots \cite{woggon:97,hawrylak:98,bimberg:98} have recently attracted enormous interest: on the one hand, they exhibit a number of atomic-like properties attributed to their zero-dimensional nature, such as ultranarrow emission peaks \cite{zrenner:94,grundmann:95} or ultralong dephasing times;~\cite{borri:01} on the other hand, the semiconductor compound gives rise to a number of novel features which lack atomic counterparts, among which multi-excitons \cite{brunner:94,landin:98,dekel:98,bayer:00a} and flexible interdot coupling \cite{schedelbeck:97,bayer:01a} are the most prominent ones. Optical excitations in semiconductors quantum dots are composed of electron-hole pairs (excitons), which become profoundly renormalized because of the resulting mutual Coulomb interactions; indeed, such Coulomb-renormalization effects have been studied exhaustively in {\em single-dot spectroscopy}~\cite{zrenner:00}\/ and are at the heart of the celebrated quantum-dot-based single-photon sources.~\cite{gerard:99,michler:00} In addition, advanced growth procedures now allow to vertically couple dots in a well-controlled manner, and to tune the coupling strength within a wide parameter range.

This flexibility renders quantum dots as ideal candidates for novel (quantum) device applications. Proposals range from cellular automata \cite{snider:99} over storage devices \cite{lundstrom:99,pazy:01} to possible registers for quantum computers.~\cite{biolatti:00} Yet, such challenging future technology requires a detailed understanding of interdot couplings and of the resulting few-particle states---issues which have only recently become subject of careful investigations. One of the crucial difficulties in these studies is the unavoidable inhomogeneous line broadening because of dot size fluctuations, inherent to any self-assembly growth procedure, which hinders the direct observation of interdot-coupling induced level splittings. Although the investigation of single quantum-dot molecules has been demonstrated and has given clear evidence of interdot coupling,~\cite{schedelbeck:97,bayer:01a} the underlying analysis faces severe problems when the change of interdot coupling is accompanied by possible variations of the lateral confinement---a delicate problem in particular for the technologically highly relevant self-assembled dots.

In this paper we present a theoretical analysis of four-wave mixing (FWM) \cite{shah:96} in an ensemble of inhomogeneously broadened coupled quantum dots, and we show that FWM spectra provide a sensitive measure of such pertinent interdot couplings. This finding rests on a number of non-trivial observations. Firstly, FWM is a technique particularly suited for the measurement of transport parameters independent of inhomogeneous broadening, e.g., exciton dephasing or biexciton binding.~\cite{borri:01,gindele:99,albrecht:96} Secondly, in the strong confinement regime the electron-hole tunneling only weakly depends on the lateral confinement,~\cite{bayer:00a,troiani.prb:02} and thus becomes independent of inhomogeneous broadening. Thirdly, as recently demonstrated~\cite{troiani.prb:02} biexciton transitions in coupled quantum dots are sensitive to interdot couplings and can be directly monitored in the optical spectra---a highly favorable finding to be contrasted with the more cumbersome situation for single-exciton transitions, where, because of symmetry, only one of the tunnel-coupled low-energy states couples to the light.~\cite{bayer:00a} Taking together all these observations, we will predict a clear-cut signature of interdot-coupling in the FWM spectra.

In our theoretical approach we start from a proto-typical level scheme for a single quantum dot consisting of:~\cite{hohenester:02,panzarini.prb:02} the groundstate $|0\rangle$ with no electron-hole pairs present; the spin-degenerate exciton states $b_\sigma^\dagger|0\rangle$ of lowest energy $\epsilon$, with $b_\sigma^\dagger$ the exciton creation operator and $\sigma$ labeling spin; the biexciton groundstate of lowest energy~\cite{hohenester:02} $b_\uparrow^\dagger b_\downarrow^\dagger|0\rangle$, whose energy $2\epsilon-\Delta$ is reduced because of Coulomb correlation effects.~\cite{dot-parameters} In Ref.~\onlinecite{troiani.prb:02} we made the important observation that in case of weak interdot couplings exciton tunneling dominates over separate electron and hole tunneling. Hence, using the above level scheme and assuming small interdot coupling throughout, we describe the double-dot system by the Hubbard-type Hamiltonian:

\begin{widetext}
\begin{equation}\label{eq:h0}
  \bm h_o(\epsilon)\cong 
    \epsilon\sum_\sigma\left(\hat n_{L\sigma}+\hat n_{R\sigma}\right)-
    t\sum_\sigma\left(b_{L\sigma}^\dagger b_{R\sigma}^{\phantom{\dagger}}+
                      b_{R\sigma}^\dagger b_{L\sigma}^{\phantom{\dagger}}
		      \right)-
    \Delta\sum_{\ell=L,R}\hat n_{\ell\uparrow}\hat n_{\ell\downarrow}
    \quad,
\end{equation}
\end{widetext}

\noindent where $L$ and $R$ denote the left and right dot, respectively, $n_{\ell\sigma}=b_{\ell\sigma}^\dagger b_{\ell\sigma}^{\phantom{\dagger}}$, and $t$ is the effective exciton-tunneling matrix element. We feel that for the purpose of our present investigation (influence of interdot coupling on FWM spectra) the use of the generic model (\ref{eq:h0}) has the advantage over the first-principles-type solution of Refs.~\onlinecite{troiani.prb:02,rontani.ssc:01} of providing deeper insight into the qualitative trends, without introducing significant differences or shortcomings. The states resulting from the solution of Eq.~(\ref{eq:h0}) are depicted in Fig.~1 as a function of the exciton interdot-coupling $t$: for linear polarization only one of the four exciton ($X$) and two of the eight biexciton states ($B_{1,2}$), respectively, couple to the light. To the lowest order of approximation, these states are associated to a symmetric superposition of excitons ($X$) and biexcitons ($B_1$) in the left and right dot, respectively, and of two excitons localized in the two spatially separated dots ($B_2$).

\begin{figure}
\includegraphics[width=0.7\columnwidth]{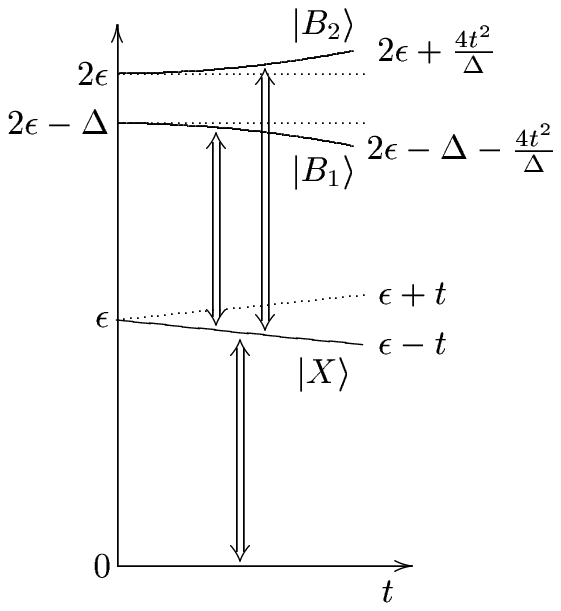}
\caption{
Schematic picture of the effective double-dot level scheme as a function of the interdot coupling strength $t$ and for $t\ll\Delta$, as obtained from the solution of Eq.~(\ref{eq:h0}). Solid lines represent the optically active exciton $|X\rangle$ and biexciton states $|B_{1,2}\rangle$, respectively; arrows `$\Longleftrightarrow$' indicate optically allowed transitions for linear polarization.~\cite{hohenester:02}
}
\end{figure}

In our simulations of the FWM spectra we assume a setup of two laser pulses with time delay $\tau$ (inset of Fig. 2),~\cite{shah:96} which propagate along directions $\bm k_1$ and $\bm k_2$. The latter pulse is diffracted by the reminiscent polarization grating produced by the first pulse, and gives rise to the FWM signal along direction $2\bm k_2-\bm k_1$. The description of the coherent (optical excitation) and incoherent (dephasing) time dynamics requires the framework of density matrices.~\cite{scully:97,haug:93} Here, the central object is the density matrix $\bm\rho$, whose time evolution is governed by the Liouville von-Neumann equation accounting for:~\cite{hohenester:02,panzarini.prb:02} the Coulomb-renormalized few-particle states; the light-coupling described within the rotating-wave and dipole approximations;~\cite{scully:97} and dephasing and relaxation due to environment interactions. In this paper we shall consider low temperatures throughout, and thus take spontaneous photon emissions as the only source of dephasing and relaxation.~\cite{borri:01} From the knowledge of $\bm\rho(t)$ we can compute the interband polarization $\bm\P(t)$, which, in turn, allows the calculation of the FWM spectra. Following Ref.~\onlinecite{banyai:95} we avoid to consider the FWM space dependency by introducing a phase shift $\phi$ between the two exciting laser pulses {\it viz.}\/ $\bm\E_o(t)\exp-i(\omega_ot-\phi)$, with $\bm\E_o$ the pulse envelope and $\omega_o$ the central pulse frequency, and compute the FWM signal according to:~\cite{banyai:95}

\begin{equation}\label{eq:fwm}
  S(t)=\int_0^{2\pi}d\phi\;\bm\P_\phi(t)e^{-i2\phi}\quad,
\end{equation}

\begin{figure}
\includegraphics[width=0.9\columnwidth]{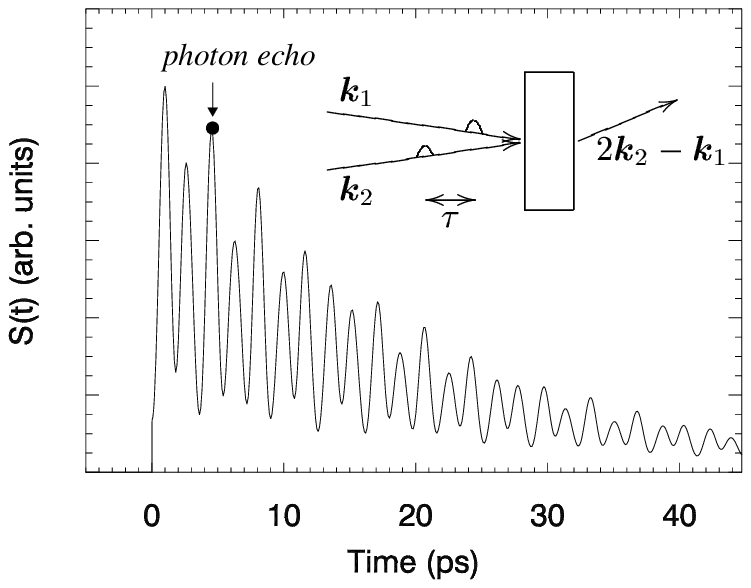}
\caption{
Results of our simulation for the transient FWM signal $S(t)$ with $\tau=5$ ps and using for clarity a short dephasing time of 20 ps;~\cite{dot-parameters} the filled circle shows the result of the corresponding simulation for an inhomogeneously broadened ensemble (full width of half maximum of 50 meV). The inset schematically depicts the proposed setup.
\vspace*{-0.5cm}
}
\end{figure}

\begin{figure}
\includegraphics[width=0.9\columnwidth]{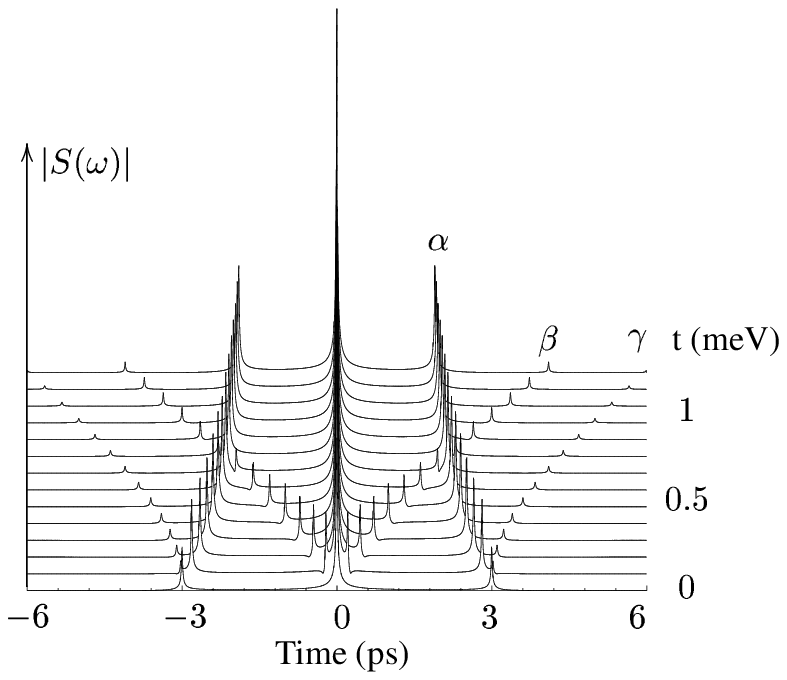}
\caption{
$|S(\omega)|$ for different values of the interdot coupling $t$ and for a dephasing time of 500 ps; spectra are offset for clarity. For discussion see text.
}
\end{figure}

\noindent where the subscript $\phi$ on $\bm\P$ is a remainder of the phase. Quite generally, in case of inhomogeneous broadening the interband polarization $\bm\P$ has to be computed for an ensemble of quantum dots with different transition energies $\epsilon$. Assuming that this broadening is much larger than the spectral width of the laser pulse, which certainly holds for typical quantum dot samples with inhomogeneous line broadenings of the order of several tens of meV, the FWM signal is given by a delta-like photon echo at time $\tau$ after the second pulse with strength $I(\tau)=\int dt\;S(t)$. However, it turns out that in case of dominant inhomogeneous broadening $I(\tau)$ is completely characterized by $S(\tau)$ for a single quantum-dot molecule;~\cite{shah:96,albrecht:96,koch:93} since computationally it is much easier to calculate the latter quantity, in the following we shall make use of this approximate description (we checked, however, its validity for a variety of time delays $\tau$).

Figure 2 shows a typical result of $S(t)$ as obtained from our simulations. We observe that $S(t)$ starts immediately after the second pulse (centered at time $t=0$), and displays a pronounced polarization beating. To gain insight into the contributing states, in Fig.~3 we plot the modulus of the Fourier transform of $S(t)$ for different values of the interdot coupling strength $t$; besides the strong signal at $\omega=0$, we observe the appearance of three peaks $\alpha$, $\beta$, and $\gamma$, which exhibit intriguing shifts with increasing $t$. Indeed, a closer analysis reveals that the peak positions can be unambiguously attributed to the $X$--$B_1$ ($\alpha$), $X$--$B_2$ ($\beta$), and $B_1$--$B_2$ ($\gamma$) transitions. It is interesting to note that while the oscillator strength of peak $\alpha$ increases with increasing $t$ because of the ``bonding'' nature of the $X$ and $B_1$ states,~\cite{troiani.prb:02} the oscillator strengths of the remaining transitions show a reversed trend because of their ``anti-binding'' nature.

To appreciate the merit of Fig.~3, we recall that our results correspond to inhomogeneous broadening of quantum-dot states (several tens of meV) {\em much larger}\/ than the relevant energy scale for interdot coupling ($\sim 1$ meV). Such coupling could not be extracted from other types of optical ensemble measurements, e.g., absorption or luminescence. Thus, FWM appears to be the ideal tool to measure coupling constants of coupled quantum-dot molecules---with the line shift of Fig.~3 providing a clearcut signature of the formation of quantum-dot molecules. We envision an experimental setup similar to Ref.~\onlinecite{borri:01} and the use of coupled-quantum-dot samples. In addition, such measurement would offer the challenging prospect of studying decoherence of entangled states in coupled dots.~\cite{bayer:00a,chen:00}

In conclusion, we have presented a theoretical study of four-wave mixing in an ensemble of inhomogeneously broadened quantum-dot molecules. We have found that the Fourier transformed spectra provide a clear signature of interdot coupling, thus rendering this technique as an ideal tool for the extraction of this pertinent parameter.

P.K. is grateful to the V\"ais\"al\"a Foundation for financial support.


\begin{thebibliography}{32}
\bibitem[{dot()}]{dot-parameters}
\bibinfo{note}{In our calculations we use a prototypical biexciton binding
  $\Delta=3$ meV. The optical dipole matrix elements for the
  groundstate-exciton and exciton-biexciton transitions are assumed to have
  equal strength;~\cite{hohenester:02,panzarini.prb:02} usual optical selection
  rules for circularly polarized light apply.}
\end{thebibliography}
\end{document}